\documentstyle[preprint,revtex]{aps}
\begin{document}
\draft
\preprint{IPT-EPFL August 1994}
\begin{title}
Impurity Spin Coupled to Strongly Correlated
Electron System:\\
Ground State Properties
\end{title}
\author{D. F. Wang and C. Gruber}
\begin{instit}
Institut de Physique Th\'eorique\\
\'Ecole Polytechnique F\'ed\'erale de Lausanne\\
PHB-Ecublens, CH-1015 Lausanne-Switzerland.
\end{instit}
\begin{abstract}
Exact results are provided for the ground state properties
of an impurity spin coupled to a conducting band of strongly
correlated electrons described by a t-J model.
The ground states are proved to be unique, apart from
the spin rotational degeneracy,
and the ground state spins are found in the following case:
(1) $J=0$, one hole hopping on a lattice in dimensions $d\ge 2$.
(2) $J\ne 0$, arbitrary number of holes hopping on a one
dimensional lattice, i.e., a Luttinger liquid interacting with
a localized impurity spin.
\end{abstract}
\pacs{PACS number: 71.30.+h, 05.30.-d, 74.65+n, 75.10.Jm }

\narrowtext

The effect of impurities Zn and Ni in the cuprate superconductors
was studied from various aspects, such as $T_c$ depression,
penetration depth depending on temperatures, conductivity, Knight shift
and nuclear relaxation rate\cite{xiao}.
Recently, Poilblanc, Scalapino and Hanke modeled the effect
of a Zn impurity with $S=0$ by a t-J model with an inert site, which
has no exchange coupling or electron transfer terms.
It was found that a hole experienced an extended effective impurity
potential, depending on the host $J/t$, and that the hole
could form bound states\cite{poil1}.

More recently, the same authors have considered an approximate
picture of the Ni impurity. The host electron system is
still described by the t-J model, and the impurity is viewed as
a hard-core repulsion plus an exchange interaction between the
impurity spin 1/2 and the host electrons\cite{poil2}.
Although in reality, Ni would have spin 1, it was argued
that the model system provides still a qualitatively
correct picture.
PSH have found that if the exchange coupling of the impurity with
the neighboring spins is ferromagnetic or weakly antiferromagnetic,
an extra hole can form bound states of different spatial
symmetries with the impurity extending to a few lattice spacings.
Exact diagonalizations through a Lanczos algorithm of small
clusters at half filling or with a single hole were
performed, for the ground state energies and dynamical correlations.
The binding was found to disappear for an antiferromagnetic coupling
exceeding $0.3J$\cite{poil2}.

Due to the strong correlation of the host electrons,
as well as due to the interaction of the impurity
with the electrons, it is very difficult
to extract some exact results for such
a complex system. Any
rigorous theorem can certainly provide helpful
informations for their correlated behaviors.
In the following, we provide some exact results
for the ground state properties of the model system
of strongly correlated electron host interacting with
a localized impurity spin, in various special cases.

The system is defined on a cubic lattice in dimensions $d$,
with Hamiltonian:
\begin{eqnarray}
H=P_G \{ J \sum_{<i,j>, i\ne j\ne 0}
( {\vec S_i}\cdot {\vec S_j} -1/4)&& n_i n_j
+t \sum_{<i,j>, i\ne j\ne 0,\sigma=\uparrow,\downarrow}
( c_{i\sigma}^\dagger c_{j\sigma} +h.c.) \nonumber\\
&&+J' \sum_{|\vec a|=1} ({\vec S_f} \cdot {\vec S_a}
-1/4) n_a \} P_G,
\end{eqnarray}
where the sum $<i,j>$ is over nearest neighbors. The first two
terms, i.e. the ordinary t-J model, represent the strongly
correlated electrons with $t>0$, the parameter associated with
the hopping matrix, and $J$ the spin exchange interaction between
the correlated electrons. The impurity spin $S_f=1/2$ is
localized at the site 0, and $J'$ is the coupling constant between
the impurity spin and the spins of the electrons at the nearest
neighboring sites. A hard core repulsion forbid any conducting
electrons to occupy the site 0. The Gutzwiller projector
$P_G$ indicates that the system is defined on the
Hilbert space with no double occupancies.
The system is rotationally invariant, the total spin
$z$ component, $S^z = \sum_{i} (n_{i\uparrow} -n_{i\downarrow})/2
+ S_{f}^z$, and the total spin of the system ${\vec S}^2$ are
constants of motions.
Since the lattice where the conducting electrons are hopping is
bipartite, we perform the unitary transformation
$U': c_{i\sigma}
\rightarrow (-1) c_{i\sigma}, c_{i\sigma}^\dagger \rightarrow
(-1) c_{i\sigma}^\dagger$ for all sites $i$ on one sublattice,
which changes the sign of $t$.
We thus consider in the following $t>0$, to which the situation
$t<0$ can be related through the unitary transformation.
Under the unitary transformation $U'$,
the total spin operator of the system remains unchanged.

We consider first a cubic lattice in dimensions $d\ge 2$ with
$N$ lattice sites ( $N$ odd ) and open boundary condition.
We want to investigate the ground states in the special case
$J=0$ and only one hole. Without the impurity spin,
the one hole case was studied by Thouless and
Nagaoka\cite{toulous,nagao,tasaki},
and it was explicitly demonstrated that the ground state is
ferromagnetic. Our goal is to show that if the
interaction between the impurity and the electrons is ferromagnetic,
i.e. $J'<0$, the ground state is completely ferromagnetic
with spin $S_G=(N-1)/2$. In other words, the ferromagnetic
coupling between the impurity spin
and the electrons preserves the ferromagnetism.
To establish this result we introduce the following basis vectors:
\begin{equation}
|\alpha>=|\{\sigma\},\sigma_f> =(-1)^{i-1} c_{1\sigma_1}^\dagger
c_{2\sigma_2}^\dagger \cdots c_{i-1\sigma_{i-1}}^\dagger
c_{i+1\sigma_{i+1}}^\dagger \cdots c_{N-1\sigma_{N-1}}^\dagger |0>
\bigotimes |\sigma_f>,
\end{equation}
where the hole is at site $i$, the spins of the electrons
are specified explicitly, and the impurity spin is $\sigma_f$.
Since each energy level has a representative in the subspace
$S^z =0$, the ground state also has
a representative in this subspace.
In the following we shall thus restrict our attention to the
subspace $S^z=0$, whose dimension is $2 \times
(N-1) \times {(N-2)! \over ({N-3\over 2})! ({N-1\over 2})!}$.
We can simply calculate the Hamiltonian matrix element
with these basis vectors:
\begin{equation}
K_{\alpha\beta} =<\alpha|H|\beta> =K_{\beta\alpha}=M_{\alpha\beta}
+m_{\alpha} \delta_{\alpha\beta},
\label{eq:condition}
\end{equation}
where the off-diagonal element $M_{\alpha\beta}$
is always non-positive, since $J'<0$ and $t>0$.
The ground state of the Hamiltonian can be expanded
as a linear combination of the basis vectors, with all the coefficients
being real numbers.
\begin{equation}
|\phi>=\sum_{\alpha} C(\alpha) |\alpha>.
\end{equation}
Using Perron-Frobenius type argument, several
interesting results were previously obtained
on ground state properties of correlated
electron systems or magnetic systems\cite{lieb1,lieb2}.
Following similar ideas, we see that the state vector
$|\bar \phi> =\sum_{\alpha} |C(\alpha)| |\alpha>$ is also a ground state
of the Hamiltonian. For a cubic lattice in dimensions $d\ge 2$, we
can carry out a simple connectivity argument to show
that all $C(\alpha)'s $ are nonzero and have the same sign.
Therefore, the ground state is unique.
Consider then the trial wavefunction
$(S^-)^{N-1\over 2} c_{1\uparrow}^\dagger
c_{2\uparrow}^\dagger\cdots c_{N-2\uparrow}^\dagger|0>\bigotimes |\uparrow>$.
We can expand $(\sum_{i=1}^L c_{i\downarrow}^\dagger c_{i\uparrow}
+S_f^-)^{N-1\over 2}$ using the property
that each object $c_{i\downarrow}^\dagger
c_{i\uparrow} $ commutes with others, without giving rise to overall
extral sign for any term in the expansion.
When the trial wavefunction is
written in terms of a linear combination of the basis vectors,
the coefficients have the same sign ( some coefficients being zero).
Therefore, the overlap between the ground state and the trial wavefunction
is nonzero,
indicating that the ground state has the same spin as the trial wavefunction,
which is $S=(N-1)/2$.
Hence, if the impurity exchange coupling is ferromagnetic $J'<0$,
we have proved that the ground state is unique (apart from
its spin degeneracy), and is
ferromagnetic with total spin $S_G=(N-1)/2$.

Let us consider the case where the impurity exchange is antiferromagnetic,
$J'>0, \, ( J=0, d\ge 2 )$. In this case, we introduce the unitary
transformation
$U=\exp[i\pi S_f^z]$. Since
\begin{eqnarray}
&&U S_f^\dagger U^\dagger =(-1) S_f^\dagger\nonumber\\
&&U S_f^- U^\dagger =(-1) S_f^-\nonumber\\
&&U S_f^z U^\dagger = S_f^z,
\end{eqnarray}
we have, with $ H'=UHU^\dagger$,
\begin{equation}
<\alpha|H'|\beta> = M_{\alpha\beta}' + m'_{\alpha\beta} \delta_{\alpha\beta},
\end{equation}
where, as before, $M'_{\alpha\beta}$ is non-positive.
Therefore, the ground state $|\phi'>$ of $H'$ is unique and
$|\phi'>=\sum_{\alpha} C(\alpha)' |\alpha>$, with all $C(\alpha)'>0$.
Consider the following trial wavefunction:
\begin{equation}
|\phi_t'>=U (\sum_{i=1}^{N-3} c_{i\downarrow}^\dagger
c_{i\uparrow})^{N-3 \over 2}
c_{1\uparrow}^\dagger c_{2\uparrow}^\dagger \cdots
c_{N-3\uparrow}^\dagger \left[ c_{N-2\uparrow}^\dagger|0>\bigotimes|\downarrow>
-c_{N-2\downarrow}^\dagger |0>\bigotimes |\uparrow> \right],
\end{equation}
which satisfies $<\phi_t'|\phi'> \ne 0$.
Since the ground state of $H'$ is unique, the same is true for
the ground state of $H$ (up to spin rotational degeneracy).
{}From the nonzero overlap between the trial
wavefunction and $|\phi'>$, it follows that
the ground state of $H, |\phi>=U^\dagger |\phi'>$,
thus has $S_G=(N-3)/2$. Loosely speaking,
the impurity forms a singlet with one electron, while the other
electrons still form a ferromagentic state due to their strong
correlation.
At this point we should
remark that the hole is not bound to the impurity spin in these
special cases. At any lattice site ( except the site 0 ), the
probability of observing the hole presence is positive, because
of the positivity property of the ground state. In this sense,
the wavefunction for the hole is extended on the lattice.

When there are many impurity spins localized on the lattice,
we can generalize the above argument in a straightforward way.
The same discussion can be carried out, if the localized
impurity spins are positioned in such a way that all the basis
vectors of the Hilbert space with fixed $S_z$
are connected to each other.
We can thus prove the
uniqueness of the ground state and find
the ground state spin.

One interesting question is what would happen if we add one more hole.
The spin background of the correlated electrons definitely affect
the motion of the holes. If the impurity is decoupled, it was
shown that the ferromagnetic state is unstable against adding
one more hole on a finite size lattice\cite{benoit,andras}.
We would expect that adding
one more hole would also destroy the ferromagnetism proved above for
this impurity system. We also wish to note that this extreme limit
$J=0$ with one hole hopping is far from the realistic situation,
where the holes are believed to hop within antiferromagnetic background for
sufficient large $J$ and sufficient number of holes doped in
real materials.

Let us finally investigate the one dimensional case.
The lattice is a closed chain of length $L$, with a
localized impurity spin at site 0.
Again, the impurity spin $\vec S_f$ interacts with the electrons on the
nearest neighboring sites. The Hamiltonian is given by:
\begin{eqnarray}
H=&&P_G \{ J \sum_{i=1}^{L-1}
( {\vec S_i}\cdot {\vec S_{i+1}} -1/4) n_i n_{i+1}
+t \sum_{\sigma} \sum_{i=1}^{L-1}
( c_{i\sigma}^\dagger c_{i+1,\sigma} +h.c.) \nonumber\\
&&+ J' (\vec S_f \cdot \vec S_{L-1} -1/4) n_{L-1}
+J' ({\vec S_f} \cdot {\vec S_1}
-1/4) n_1 \}  P_G.
\end{eqnarray}
We consider now the general case $J>0, t>0$.
This model represents a Luttinger liquid coupled to a
localized impurity spin, which is different from
the ordinary Kondo impurity model, where a Fermi liquid
is interacting with a localized impurity spin.

Before we carry out further steps, we perform the
unitary transformation $U'$ to change the sign of the
hopping matrix element.
As before we first consider the situation where the impurity
coupling is ferromagnetic $J'<0$. The number of the electrons
$N_e$ is assumed to be odd, and we work in the subspace where
$S^z$, the z-component of the total spin of the system, is
zero. The basis vectors of the Hilbert space can be choosen
in the following way:
\begin{eqnarray}
&&| x_1,\cdots,x_{N_e-M}; y_1,\cdots,y_M;\uparrow>\nonumber\\
&&=(-1)^{M(M-1)/2} c_{x_1\uparrow}^\dagger \cdots
c_{x_{N_e-M}\uparrow}^\dagger c_{y_1\downarrow}^\dagger \cdots
c_{y_M\downarrow}^\dagger |0> \bigotimes |\uparrow>_f,
\nonumber\\
&&|x_1,\cdots,x_{N_e-M+1}; y_1,\cdots,y_{M-1};\downarrow>\nonumber\\
&&=(-1)^{(M-1)(M-2)/2} c_{x_1\uparrow}^\dagger\cdots
c_{x_{N_e-M+1}\uparrow}^\dagger c_{y_1\downarrow}^\dagger \cdots
c_{y_{M-1}\downarrow}^\dagger |0>\bigotimes |\downarrow>_f,
\end{eqnarray}
where $x_1<x_2<\cdots<x_{N_e-M+1}, y_1<y_2<\cdots<y_M$, $N_e$ is
the number of the electrons on the lattice, and $N_e-M+1=M$ such that
$S^z=0$.

When $J'<0$, choosing the basis vectors in the above
fashion, we see that the off-diagonal matrix elements of the Hamiltonian
are always non-positive. Since $J\ne 0$,
all the basis vectors are connected
to each other, and therefore the ground state
is unique and a positive vector.
Moreover, the ground state has spin $S_G=1$,
since it has nonzero overlap with the trial wavefunction:
\begin{eqnarray}
&&|trial>=(c_{1\uparrow}^\dagger c_{2\downarrow}^\dagger -
c_{1\downarrow}^\dagger c_{2\uparrow}^\dagger)
(c_{3\uparrow}^\dagger c_{4\downarrow}^\dagger -
c_{3\downarrow}^\dagger c_{4\uparrow}^\dagger)
\cdots\nonumber\\
&&\cdots(c_{N_e-2\uparrow}^\dagger c_{N_e-1\downarrow}^\dagger -
c_{N_e-2\downarrow}^\dagger c_{N_e-1\uparrow}^\dagger)
\left[c_{N_e\uparrow}^\dagger|0>\bigotimes |\downarrow>
+c_{N_e\downarrow}^\dagger |0> \bigotimes |\uparrow> \right],
\end{eqnarray}
which is an eigenstate of total spin operator with eigenvalue $S=1$.

When $J'>0$, that is, the coupling between the impurity
spin and the conducting band is antiferromagnetic, we perform the
unitary transformation $U=\exp{i\pi S_f^z}$ as before.
After the operation, we can choose the basis vectors in a similar fashion
for the transformed Hamiltonian, and the off-diagonal matrix elements
of this Hamiltonian are non-positive. The connectivity argument shows
that its ground state is unique, indicating that the ground state of the
original Hamiltonian is also unique. This ground state spin can be found
to be $S_G=0$.

In the one hole case,
as the ground state is a positive vector, the probability of
finding the hole at any lattice site is nonzero, indicating that the hole
is not bound to the impurity spin. When there are many holes on
the chain, it is also true that
not a single hole can be bound to the impurity spin.

In summary, we have obtained the ground state spins for the system
where an impurity spin interacts with strongly correlated electrons
in various special cases. In particular, the model Hamiltonian
in one dimension represents a Luttinger liquid coupled to
a localized impurity spin. It remains to find how the impurity
spin behaves under the influence of the Luttinger liquid,
such as how much the impurity spin
is screened, whether a dynamical scale is generated as in
the usual Kondo model, where a Fermi liquid is interacting with
a localized impurity spin\cite{anderson,andrei1,andrei2}.

We would like to thank Dr. N. Macris for conversations.
This work was supported by the World Laboratory.

\end{document}